\documentclass[sigconf]{acmart}

\usepackage{booktabs} 
\usepackage{multirow} 
\usepackage[ruled,vlined]{algorithm2e}

\setcopyright{rightsretained}

\acmConference[ESEM2018]{12th International Symposium on Empirical Software Engineering and Measurement}{October 2018}{Oulu, Finland}
\acmYear{2018}

\begin{document}
\title{An empirical study of public data quality problems in cross project defect prediction}

\author{Zhongbin Sun, Junqi Li, Heli Sun}
\affiliation{%
  \institution{School of Electronic and Information Engineering, Xi'an Jiaotong University}
  \city{Xi'an}
  \state{Shaanxi}
  \postcode{710049}
  \country{China}
}
\email{zhongbin725@mail.xjtu.edu.cn, vivi528@stu.xjtu.edu.cn, hlsun@xjtu.edu.cn}


\begin{abstract}
Background: Two public defect data, including Jureczko and NASA datasets, have been widely used in cross project defect prediction (CPDP). The quality of defect data have been reported as an important factor influencing the defect prediction performance and Shepperd et al. have researched the data quality problems in NASA datasets. However, up to now, there is no research focusing on the quality problems of Jureczko datasets which are most widely used in CPDP. Aims: In this paper, we intend to investigate the problems of identical and inconsistent cases in Jureczko datasets and validate whether removing these problematic cases will make a difference to defect prediction performance in CPDP. Method: The problems of identical and inconsistent cases are reported from two aspects, respectively in each individual dataset and in a pair of datasets from different releases of a software project. Then a cleaned version of Jureczko datasets is provided by removing duplicate and inconsistent cases. Finally three training data selection methods are employed to compare the defect prediction performance of cleaned datasets with that of original datasets. Results: The experimental results in terms of AUC and F-Measure show that most datasets obtain very different defect prediction performance. Conclusions: It is very necessary to study the data quality problems in CPDP and the cleaned Jureczko datasets may provide more reliable defect prediction performance in CPDP.
   
\end{abstract}


\maketitle

\section{Introduction}
Software systems, ubiquitous in the real world, have been widely used in different applications. With the continuous increase of software size and complexity, software quality assurance has become increasingly important. One way to improve software quality is software defect prediction, which has been an important research topic in the software engineering field for more than 40 years. Software defect prediction aims at finding the defect-prone modules in software and thus helps organizations to allocate the limited resources reasonably, which is an efficient means to relieve the effort in software code inspection or testing. Currently within project defect prediction (WPDP) \cite{Menzies2007a, Lessmann2008benchmarking, Song2011A, Sun2012Using, Bennin2017MAHAKIL} and cross project defect prediction (CPDP) \cite{Briand2002Assessing, Zimmermann2009Cross-project, Xia2016HYDRA, Zhang2016Cross-project, Herbold2017Comparative} are two popular but different directions of defect prediction research.

 In the field of software defect prediction research, most approaches employ machine learning classifiers to build a prediction model from training data mined from the software repositories, and the model is used to identify software defects in the test data. In WPDP, the training data and the test data come from the same project and thus for the purpose of ensuring the efficiency of WPDP, there are two essential assumptions, namely (1) the training data and the test data should be independently and identically distributed and (2) there are enough historical defect data to build a good prediction model. However, in practice, there are difficulties in collecting and organizing the defect data for many companies and the defect data are usually absent. In addition, the new projects often do not have enough defect data to build a prediction model. Therefore the traditional WPDP is usually inefficient with the scarcity of training data and many software engineering researchers have focused their research on CPDP in recent years. 
 
 Instead of using historical defect data from the same project as the training data, the CPDP methods aim to build the prediction model on one project (known as source) with sufficient historical data and use the model to make prediction on the other project (known as target). Specially, Herbold et al. \cite{Herbold2017Global} have categorized the CPDP research into mixed CPDP and strict CPDP according to the difference of source data. In mixed CPDP \cite{Turhan2011Empirical, He2012investigation, Turhan2013Empirical}, the source data are from old revisions of the same project together with data from other projects while the strict CPDP \cite{Turhan2009relative, Peters2013Better, Yi2017Training, Zhang2017Cross} research only employ data from other projects as the source data. In \cite{Turhan2009relative}, Turhan et al. report that the CPDP model would have a low prediction performance with all the available source project data and they find that applying the nearest neighbor filtering to the source project data could considerably improve the prediction performance of the resulting CPDP model. Therefore, in the last few years, many studies have focused on the relevance filters \cite{Bettenburg2012Think, Menzies2013Local, Peters2013Better, Herbold2013Training, Bettenburg2015Towards, Kawata2016Improving, Li2017Evaluating, Yi2017Training}, which aims to employ different approaches to find the appropriate training data for building CPDP models. 
 
Hosseini et al. \cite{Hosseini2017A} report that Jureczko \cite{Jureczko2010Using, Jureczko2010Towards} and NASA \cite{Menzies2007a} are the most widely used datasets in current CPDP studies, which occupies the proportion of 65\% and 54\% respectively. In fact, Shepperd et al.  \cite{Shepperd2013Data} have reported the data quality problems of NASA and provided two cleaned versions of the NASA data, namely $DS'$ and $DS''$. The difference between $DS'$ and $DS''$ is that whether the duplicate and inconsistent cases are removed and Rodriguez et al. \cite{Rodriguez2014Preliminary} find that there are obvious performance differences between the two datasets for WPDP. However, to our best knowledge, no CPDP research mentioned above have realized the data quality problems in the Jureczko data sets. The data quality in CPDP is a crucial problem and it may greatly influence the prediction performance, especially for the relevancy filter-based training data selection methods. 

In present study we have firstly investigated the data quality problems of Jureczko datasets, including the identical and inconsistent cases in individual datasets and in a pair of datasets from different releases of a software project. Then a data cleaning method is proposed to remove the duplicate and inconsistent cases in the Jureczko datasets and a cleaned version of Jureczko datasets is provided. Finally three popular training data selection methods ( Global Filter \cite{Menzies2013Local}, Burak Filter \cite{Turhan2009relative} and Peters Filter \cite{Peters2013Better}) are selected to compare the prediction performance of the cleaned datasets with that of original datasets. The experimental results in terms of AUC and F-Measure show that most problematic datasets obtain very different prediction performance, which indicates the effectiveness and necessity of our data cleaning approach for the Jureczko datasets.

The rest of this paper is organized as follows. Section 2 introduces the related work in cross project defect prediction. Section 3 provides a detailed description of Jureczko datasets and investigates the data quality problems in the Jureczko datasets. We provide a data cleaning method as well as the cleaned version of Jureczko datasets in Section 4. Section 5 describes the comparative experiments between the cleaned datasets and original datasets, and then reports the corresponding results analysis. Finally Section 6 concludes this paper.  

\section{Related Work}
In recent years, CPDP has drawn wide attention from software engineering researchers. A wide range of topics including verification of CPDP in different programming languages, different granularities and different companies, have been studied in existing CPDP research \cite{Briand2002Assessing, Watanabe2008Adapting, Turhan2009relative, Zimmermann2009Cross-project, He2012investigation, Menzies2013Local, Peters2013Better, Peters2013Balancing, Zhang2017Data, Yu2017feature, Wu2017Cross-project, Jing2017improved}. These studies have greatly improved our understanding of the practical value of cross project defect prediction. It seems that CPDP has become an active research area in the field of defect prediction and has achieved promising defect prediction results \cite{He2012investigation, Nam2015Heterogeneous, Hosseini2016Search}. However, Herbold et al. \cite{Herbold2017Comparative} and Hosseini et al. \cite{Hosseini2017A} draw the conclusion that CPDP is still a challenge and requires more research before trustworthy application in practice. 

Many CPDP work mainly focus on verifying the feasibility of the CPDP methods \cite{ Rahman2012Recalling, Nam2013Transfer, Zhang2016Towards, Ma2012Transfer, Herbold2017systematic}. Briand et al.\cite{Briand2002Assessing} firstly propose cross project defect prediction, using the open source software Xpose defect data to make predictions for Jwriter.  Carmago et al. \cite{Camargo2009Towards} found that making the underlying distributions between the source data and target data similar may increase the quality of defect prediction. Nam et al. \cite{Nam2013Transfer} extend a state-of-the-art transfer learning method TCA and propose a method TCA+. They found that TCA+ could significantly improve cross project prediction performance. In the case that different projects may provide different metrics, Nam et al. \cite{Nam2015Heterogeneous} present the heterogeneous defect prediction approach based on transfering knowledge. For solving the class imbalance problem in CPDP, Jing et al. \cite{Jing2017improved} has proposed an improved SDA based defect prediction framework and it could significantly outperforms related methods. Zimmermann et al. \cite{Zimmermann2009Cross-project} study cross project defect prediction models on a large scale and they investigate three factors that may influence the success of CPDP for the first time, including data, domain and process. They found that the success rate of the CPDP is very low (only 3.4\% of 622 models are successful). In addition, Zimmermann et al. also found that the CPDP between the projects is not symmetrical.

The results obtained in \cite{Zimmermann2009Cross-project} has shown the importance of training data selection. Turhan et al.\cite{Turhan2013Empirical} found that blindly selecting training data for CPDP can easily lead to high false negative rates for the prediction results. In addition, Hosseini et al. \cite{Hosseini2016Search}  have shown that the selection of training data can lead to better performance in cross project defect prediction. Therefore many researchers have focused their research on how to select appropriate training data for the target project and there are two lines of training data selection, including source projects selection and instances selection. 

In the research of source projects selection, He et al.\cite{He2012investigation} found that if a best possible training set of three software projects from a set of available projects is selected through a strategy with a posteriori knowledge, a defect prediction success rate of over 50\% may be achieved. Then they propose a strategy for the selection of source projects with decision trees over the distributional characteristics of training data and test data. However, their method does not scale with the number of datasets and the runtime of their method is exponential. Therefore based on the assumption that similar distributions lead to better results for cross project defect prediction, Herbold et al.\cite{Herbold2013Training} propose two distance-based strategies for the selection of source projects by using  the distributional characteristics of available data. In addition, Khoshgoftaar et al. \cite{Khoshgoftaar2009Software} propose to use a combination of multiple classifiers and data from multiple products for CPDP. They demonstrated that the combination of predictions of multiple learners trained on multiple datasets could improve the prediction performance of a learner induced on a single dataset.  Aarti et al.\cite{Aarti2017Effect} investigated the prediction accuracy of cross project defect prediction and indicate that cross project prediction can provide better prediction accuracy by combining different source projects. 

\begin{table*}[!htbp]
	\caption{A statistic summary of the original Jureczko datasets}
	\label{OrigData}
	\small	
	\begin{tabular}{ccccc|ccccc}
		\toprule
		ID & Dataset & \#Case & \#Defective & \%Defective & ID & Dataset& \#Case& \#Defective &\%Defective\\
		\hline
		1 &  ant1.7 &   745 & 166 &  22\% &34 & poi3.0 & 442 &  281 & 64\% \\
		
		2 &  arc &  234 &  27 & 12\% &	35 & prop1 & 18471 & 2738 & 15\% \\
		
		3 & berek &  43 & 16 & 37\% &36 & prop2 &  23014 & 2431 &   11\% \\
		
		4 & camel1.0 &  339 &  13 &  4\% &37 & prop3 & 10274 &  1180 & 11\% \\
		
		5 & camel1.2 &  608 &  216 & 36\% &38 &  prop4 & 8718 &  840 &  10\% \\
		
		6 & camel1.4 & 872 &  145 & 17\% &39 & prop5 & 8516 &  1299 &  15\% \\
		
		7 & camel1.6 &  965 &  188 & 19\% &40 & prop6 &   660 &  66 &  10\% \\
		
		8 & ckjm &   10 &    5 & 50\% &41 &   redaktor &  176 &  27 &  15\% \\
		
		9 &  elearning &  64 &  5 & 8\% &42 &   serapion &  45 &  9 & 20\% \\
		
		10 & forrest0.6 & 6 &   1 & 17\% &	43 &  skarbonka &  45 &   9 & 20\% \\
		
		11 & forrest0.7 &   29 &   5 & 17\% &44 &   sklebagd &  20 & 12 &   60\% \\
		
		12 & forrest0.8 &  32 & 2 & 6\% &45 & synapse1.0 & 157 & 16 & 10\% \\
		
		13 & intercafe & 27 & 4 &  15\% & 46 & synapse1.1 & 222 & 60 &  27\% \\
		
		14 &  ivy1.1 &  111 & 63 &  57\% &47 & synapse1.2 &  256 & 86 &  34\% \\
		
		15 &  ivy1.4 &  241 & 16 & 7\% & 48 & systemdata & 65 &  9 &  14\% \\
		
		16 &  ivy2.0 & 352 & 40 &  11\% &	49 & szybkafucha &  25 &   14 &  56\% \\
		
		17 &   jedit3.2 & 272 & 90 & 33\% &	50 & termoproject & 42 & 13 & 31\% \\
		
		18 &   jedit4.0 &  306 &  75 &  25\% &	51 & tomcat &  858 & 77 &  9\% \\
		
		19 & jedit4.1 & 312 & 79 &  25\% &	52 & velocity1.4 & 196 & 147 & 75\% \\
		
		20 & jedit4.2 & 367 &  48 &  13\% &53 & velocity1.5 &  214 & 142 & 66\% \\
		
		21 & jedit4.3 & 492 &  11 & 2\% & 54 & velocity1.6 & 229 &  78 &  34\% \\
		
		22 &  kalkulator & 27 &  6 & 22\%& 55 & workflow &  39 &  20 &  51\%  \\
		
		23 &  log4j1.0 & 135 & 34 &  25\% &	56 & wspomaganiepi & 18 & 12 & 67\% \\
		
		24 &  log4j1.1 &  109 & 37 & 34\% &57 &  xalan2.4 & 723 & 110 & 15\% \\
		
		25 &  log4j1.2 &  205 & 189 &  92\% &58 & xalan2.5 &  803 &  387 & 48\% \\
		
		26 &  lucene2.0 & 195 &  91 &  47\% & 59 & xalan2.6 &  885 & 411 &  46\% \\
		
		27 &  lucene2.2 & 247 & 144 & 58\% &60 &  xalan2.7 & 909 & 898 & 99\% \\
		
		28 &  lucene2.4 & 340 & 203 & 60\% &61 & xerces1.2 & 440 & 71 & 16\% \\
		
		29 &  nieruchomosci & 27 & 10 & 37\% & 62 & xerces1.3 & 453 & 69 & 15\% \\
		
		30 & pdftranslator & 33 & 15 & 45\% &63 & xerces1.4 & 588 & 437 & 74\% \\
		
		31 &  poi1.5 &   237 &   141 &  59\% &64 & xercesinit & 162 & 77 & 48\% \\
		
		32 &  poi2.0 &  314 &  37 &12\% &	65 & zuzel &  29 &  13 & 45\% \\
		
		33 &  poi2.5 &   385 &  248 &  64\% &  &  &  &  &\\				
		\bottomrule
	\end{tabular}			
\end{table*}

Turhan et al.\cite{Turhan2009relative} have employed a nearest neighbor filtering method to select appropriate  instances for building the CPDP model as they found that the predictive performance would be lower if all available source project data were used. The nearest neighbor filtering method use test instances to guide the selection of training data. After applying the nearest neighbor filtering on source project data, Turhan et al. found that the resulting predictive performance of the CPDP model is significantly improved. Bettenburg et al. \cite{Bettenburg2012Think} propose a local cluster guided selection method with a popular clustering algorithm.  Then based on the filter proposed in \cite{Bettenburg2012Think}, Menzies et al.\cite{Menzies2013Local} proposes a neighbor cluster guided selection method. Kawata et al.\cite{Kawata2016Improving} introduce a density-based spatial clustering to guide training data filtering. Moreover, Peters et al. \cite{Peters2013Better} propose a training instance guided selection method with k-means clustering algorithm. In this method, the training data and the test data are combined to a dataset and then k-means clustering algorithm is used to obtain different clusters. The clusters that contain at least one test instance are kept and the closest test instance in the same cluster for each training instance is found to label corresponding training instance. Finally for the test instances, the Euclidean distance is used to select the nearest training case in the training cases that are labelled with the corresponding test case. All the selected training cases are combined into the training data for building a prediction model.

The Jureczko datasets \cite{Jureczko2010Using, Jureczko2010Towards} have been widely used in the research mentioned above and Hosseini et al. \cite{Hosseini2017A} have reported that Jureczko datasets are the most widely used datasets in current CPDP studies. However, to the best of our knowledge, no previous studies have noticed the data quality problems in Jureczko datasets. Shepperd et al.  \cite{Shepperd2013Data} have reported the data quality problems of NASA and provided two cleaned versions of the NASA data, in which the difference is whether removing duplicate and inconsistent cases or not. Then Rodriguez et al. \cite{Rodriguez2014Preliminary} found that the two versions of NASA data could achieve obvious defect prediction performance differences. Therefore it is urgent and necessary to investigate the data quality problems in the Jureczko datasets.

\section{Investigation}

\subsection{Jureczko Datasets}
The most widely used Jureczko datasets are parts of the PROMISE repository \footnote{The data is publicly available online: http://openscience.us/repo/defect/ck/} and have been provided by Jureczko et al. \cite{Jureczko2010Using, Jureczko2010Towards}. It consists of 33 different open source software development projects. However, until now the PROMISE repository has provided the metrics and defect of 32 projects except the \emph{pbeans} project. Therefore hereinafter we will investigate the obtained 32 projects.

In present study, the Jureczko datasets consist of 65 releases of 32 different software projects. In an individual dataset, each case represents a java class of the corresponding release and it contains two parts: features including 20 static code metrics and a labeled feature indicating the number of defects in that class. In this study, we preprocess the Jureczko datasets by considering a class as defective if the value of the labeled feature is equal or greater than 1. Therefore the obtained Jureczko datasets with 20 features and a binary label (defective or defect-free) are considered as the original datasets for future investigation. Table \ref{OrigData} provides a brief summary of the original Jureczko datasets, including the number of cases (\#Case), the number of defective cases (\#Defective) and the percentage of the defective cases (\%Defective). 


\subsection{Data Quality Problems}
Rodriguez et al. \cite{Rodriguez2014Preliminary} have found that removing the duplicate and inconsistent cases in NASA datasets may have a significant impact on the prediction performance of WPDP. Thus in this section, we will investigate the problems of identical and inconsistent cases in Jureczko datasets. Apart from the data quality problems in each individual dataset, the data quality problems between different releases of a particular software project are also studied as the mixed CPDP uses the old releases of the target project for building the prediction model. In the following, the description of identical cases and inconsistent cases are firstly introduced.

\textbf{Identical Cases:} Two or more cases contain identical values for all features as well as the class label.

\textbf{Inconsistent Cases:} Two or more cases contain identical values for all features but the class labels differ.

Table \ref{WithinQuality} provide the detailed data quality analysis for each individual Jureczko datasets, including the number of identical cases (\#Inc) and the number of inconsistent cases (\#Ide).
\begin{table}[!h]
	\caption{Detailed data quality analysis for each dataset }
	\label{WithinQuality}
	\small
	\begin{tabular}{ccc|ccc}
		\toprule
	Dataset & \#Inc & \#Ide &Dataset& \#Inc & \#Ide\\
		\hline	
		ant1.7 & 0 & 36 & poi3.0 & 9 & 70 \\
		
		arc & 2 & 28 & prop1 & 10399 & 11860 \\
		
		berek & 0 & 0 & prop2 & 3483 & 13527 \\
		
		camel1.0 & 0 & 22 & prop3 & 2645 & 7734 \\
		
		camel1.2 & 38 & 58 & prop4 & 2550 & 6577 \\
		
		camel1.4 & 0 & 101 & prop5 & 1687 & 5669 \\
		
		camel1.6 & 4 & 117 & prop6 & 47 & 372 \\
		
		ckjm & 0 & 0 & redaktor & 1 & 10 \\
		
		elearning & 0 & 12 & serapion & 0 & 2 \\
		
		forrest0.6 & 0 & 0 & skarbonka & 0 & 0 \\
		
		forrest0.7 & 0 & 2 & sklebagd & 0 & 0 \\
		
		forrest0.8 & 0 & 2 & synapse1.0 & 0 & 6 \\
		
		intercafe & 0 & 0 & synapse1.1 & 1 & 12 \\
		
		ivy1.1 & 1 & 4 & synapse1.2 & 0 & 18 \\
		
		ivy1.4 & 0 & 10 & systemdata & 0 & 4 \\
		
		ivy2.0 & 0 & 14 & szybkafucha & 0 & 0 \\
		
		jedit3.2 & 0 & 8 & termoproject & 0 & 2 \\
		
		jedit4.0 & 1 & 6 & tomcat & 0 & 98 \\
		
		jedit4.1 & 0 & 8 & velocity1.4 & 0 & 28 \\
		
		jedit4.2 & 0 & 8 & velocity1.5 & 0 & 27 \\
		
		jedit4.3 & 32 & 24 & velocity1.6 & 0 & 31 \\
		
		kalkulator & 1 & 0 & workflow & 0 & 2 \\
		
		log4j1.0 & 0 & 0 & wspomaganiepi & 0 & 0 \\
		
		log4j1.1 & 0 & 0 & xalan2.4 & 0 & 49 \\
		
		log4j1.2 & 0 & 6 & xalan2.5 & 3 & 86 \\
		
		lucene2.0 & 1 & 4 & xalan2.6 & 39 & 189 \\
		
		lucene2.2 & 2 & 6 & xalan2.7 & 0 & 208 \\
		
		lucene2.4 & 0 & 6 & xerces1.2 & 34 & 114 \\
		
		nieruchomosci & 0 & 2 & xerces1.3 & 1 & 118 \\
		
		pdftranslator & 0 & 0 & xerces1.4 & 5 & 132 \\
		
		poi1.5 & 19 & 31 & xercesinit & 3 & 19 \\
		
		poi2.0 & 1 & 48 & zuzel & 1 & 0 \\
		
		poi2.5 & 2 & 50 &   &   &   \\
		\bottomrule
	\end{tabular}  
\end{table}


From Table \ref{WithinQuality}, it could be observed that 52 datasets suffer from the problem of identical cases and 28 datasets suffer from the problem of inconsistent cases. The identical cases in the corresponding 52 datasets ranges from 1.76\% (lucene2.4) to 75.44\% (prob1). Likewise, the inconsistent cases in the 28 datasets take a proportion of range from 0.22\% (xerces1.3) to 56.3\% (prob1). Particularly, some datasets are severely affected by both the problems of inconsistent cases and identical cases, such as prop1, prop3 and prob4. 

Table \ref{CompareQuality} provide the detailed data quality analysis for two different releases of a same project. Note that in Table \ref{OrigData} there is only one release in some projects and thus these projects with one release are absent in Table \ref{CompareQuality}.

\begin{table*}[!h]
	\caption{Detailed data quality analysis for different releases}
	\label{CompareQuality}
	\small 
	\begin{tabular}{cccc|cccc}
		\toprule
		Release1 & Release2 & \#Identical & \#Inconsistent& Release1 & Release2& \#Identical& \#Inconsistent\\
		\hline	
		
		camel1.0 & camel1.2 & 112 & 19 & prop1 & prop2 & 17748 & 4940 \\
		
		camel1.0 & camel1.4 & 111 & 2 & prop1 & prop3 & 12010 & 3492 \\
		
		camel1.0 & camel1.6 & 92 & 3 & prop1 & prop4 & 19283 & 5377 \\
		
		camel1.2 & camel1.4 & 408 & 52 & prop1 & prop5 & 10795 & 3312 \\
		
		camel1.2 & camel1.6 & 406 & 39 & prop1 & prop6 & 610 & 28 \\
		
		camel1.4 & camel1.6 & 863 & 46 & prop2 & prop3 & 10591 & 2101 \\
		
		forrest0.6 & forrest0.7 & 1 & 0 & prop2 & prop4 & 14904 & 3469 \\
		
		forrest0.6 & forrest0.8 & 1 & 0 & prop2 & prop5 & 7934 & 1848 \\
		
		forrest0.7 & forrest0.8 & 18 & 0 & prop2 & prop6 & 107 & 0 \\
		
		ivy1.1 & ivy1.4 & 9 & 1 & prop3 & prop4 & 7511 & 1512 \\
		
		ivy1.1 & ivy2.0 & 3 & 0 & prop3 & prop5 & 5327 & 176 \\
		
		ivy1.4 & ivy2.0 & 31 & 0 & prop3 & prop6 & 156 & 3 \\
		
		jedit3.2 & jedit4.0 & 71 & 4 & prop4 & prop5 & 8079 & 1432 \\
		
		jedit3.2 & jedit4.1 & 58 & 0 & prop4 & prop6 & 33 & 0 \\
		
		jedit3.2 & jedit4.2 & 31 & 0 & prop5 & prop6 & 30 & 6 \\
		
		jedit3.2 & jedit4.3 & 8 & 0 & synapse1.0 & synapse1.1 & 24 & 6 \\
		
		jedit4.0 & jedit4.1 & 67 & 4 & synapse1.0 & synapse1.2 & 9 & 1 \\
		
		jedit4.0 & jedit4.2 & 36 & 0 & synapse1.1 & synapse1.2 & 82 & 10 \\
		
		jedit4.0 & jedit4.3 & 8 & 0 & velocity1.4 & velocity1.5 & 7 & 17 \\
		
		jedit4.1 & jedit4.2 & 68 & 1 & velocity1.4 & velocity1.6 & 2 & 15 \\
		
		jedit4.1 & jedit4.3 & 20 & 0 & velocity1.5 & velocity1.6 & 44 & 48 \\
		
		jedit4.2 & jedit4.3 & 66 & 0 & xalan2.4 & xalan2.5 & 211 & 147 \\
		
		log4j1.0 & log4j1.1 & 46 & 5 & xalan2.4 & xalan2.6 & 233 & 68 \\
		
		log4j1.0 & log4j1.2 & 7 & 42 & xalan2.4 & xalan2.7 & 22 & 95 \\
		
		log4j1.1 & log4j1.2 & 5 & 42 & xalan2.5 & xalan2.6 & 417 & 204 \\
		
		lucene2.0 & lucene2.2 & 40 & 24 & xalan2.5 & xalan2.7 & 93 & 169 \\
		
		lucene2.0 & lucene2.4 & 12 & 8 & xalan2.6 & xalan2.7 & 107 & 664 \\
		
		lucene2.2 & lucene2.4 & 30 & 26 & xerces1.2 & xerces1.3 & 871 & 112 \\
		
		poi1.5 & poi2.0 & 93 & 36 & xerces1.2 & xerces1.4 & 347 & 592 \\
		
		poi1.5 & poi2.5 & 92 & 37 & xerces1.2 & xercesinit & 50 & 49 \\
		
		poi1.5 & poi3.0 & 41 & 23 & xerces1.3 & xerces1.4 & 435 & 516 \\
		
		poi2.0 & poi2.5 & 117 & 266 & xerces1.3 & xercesinit & 17 & 80 \\
		
		poi2.0 & poi3.0 & 68 & 26 & xerces1.4 & xercesinit & 14 & 81 \\
		
		poi2.5 & poi3.0 & 68 & 47 &   &   &   &   \\
		\bottomrule	
	\end{tabular}  
\end{table*}


From Table \ref{CompareQuality}, it could be observed that all the pairs of different releases for a particular software project suffer from the quality problem of identical cases and the number of identical cases range from 1 to 19283. In addition, most of the pairs of different releases also suffer from the problem of inconsistent cases and the number of inconsistent cases range from 1 to 5377. 

In a word, there are many identical and inconsistent cases not only in each individual dataset but also in different releases of a particular software project. It is reasonbale to doubt that whether so many identical and inconsistent cases may have unknowable effect on the prediction performance of cross project defect prediction. Therefore for the purpose of validating whether such effect is real, the original Jureczko datasets are cleaned in Section \ref {CleaningSection} and we conduct comparative experiments using the original and cleaned Jureczko datasets in Section \ref{ExperimentSection}.     

\section{Data Cleaning} \label{CleaningSection}

In this section, a data cleaning method is firstly proposed to deal with the problematic cases in the  Jureczko datasets as there are many identical and inconsistent cases found in the Jureczko datasets. Then  
the cleaned datasets are provided as well as a brief statistic summary of the cleaned datasets.

Algorithm \ref{DataCleaning} provides the data cleaning method for Jureczko datasets in details. Particularly, the method consists of two main steps: remove duplicate cases and remove inconsistent cases. In Algorithm \ref{DataCleaning}, lines 6-9 deal with the identical cases and remove the subsequent cases duplicate with the current case. The identical cases could be recognized when the features and class label of two cases are same. In addition, lines 10-13 are used to remove the two inconsistent cases which have the identical feature values but different class label values. Note that the order of the two steps can not be swapped, other some inconsistent cases may not be removed. Shepperd et al. \cite{Shepperd2013Data} has provided a simple example for explaining why these two steps could not be swapped. 
\begin{algorithm}[!h]
	\caption{Data Cleaning Method}
	\label{DataCleaning}
	\LinesNumbered
	\small
	\KwIn{$Data = \{DS_1,DS_2,\cdots, DS_k\}$ //The original datasets}
	\KwOut{$NewData = \{NDS_1, NDS_2, \cdots, NDS_k\}$ // The cleaned datasets}
	// M - the number of cases in DS \\
	// N - the number of features (including class label) in DS \\
	// DS.Value[i][j] - the value of feature j in case i in DS\\
	
	\For{each $DS_k\in Data$}{
		$NDS_k = DS_k$;
		
		\For( //Step 1: remove duplicate cases){$i=1$ to $M-1$} {
			\For{$j=i+1$ to $M$}{  
			  \If{$NDS_k.Value[i][1...N] \equiv  NDS_k.Value[j][1...N]$}
			  {
			 	$NDS_k = NDS_k - NDS_k.Value[j][1...N]$;
			  }
			}    
		}
		\For( //Step 2: remove inconsistent cases){$i=1$ to $M-1$} {
			\For{$j=i+1$ to $M$}{  
					\If{$NDS_k.Value[i][1...N-1] \equiv NDS_k.Value[j][1...N-1]$ and $NDS_k.Value[i][N] \ne NDS_k.Value[j][N]$}{
						$NDS_k = NDS_k - NDS_k.Value[i][1...N]$;
						$NDS_k = NDS_k - NDS_k.Value[j][1...N]$;
					}
			}    
		}
	}
return $NewData = \{NDS_1, NDS_2, \cdots, NDS_k\}$; 
\end{algorithm}

By applying the data cleaning method Algorithm \ref{DataCleaning} to Jureczko datasets, a cleaned version of the datasets could be obtained and it is available from \textbf{\url{http://gr.xjtu.edu.cn/web/zhongbin725/datasets}}. Then for the purpose of showing the difference between the original data and cleaned data, Table \ref{NewData} provides a brief statistic summary of the cleaned Jureczko datasets. Note that different from Table \ref{OrigData}, Table \ref{NewData} shows not only the number of cases (\#Case) and defective cases (\#Defective) but also the number of deleted cases (\#delCase) and deleted defective cases (\#delDefective).

\begin{table*}
	\caption{A statistic summary of the cleaned Jureczko datasets}
	\label{NewData}
	\small
	\begin{tabular}{cccccc|cccccc}
		\toprule
	ID &Dataset & \#Case & \#delCase & \#Defective& \#delDefective & ID & Dataset & \#Case & \#delCase & \#Defective& \#delDefective \\
		\hline
		1 & ant1.7 & 724 & 21 & 166 & 0 & 34 & poi3.0 & 398 & 44 & 255 & 26 \\
		
		2 & arc & 213 & 21 & 25 & 2 & 35 & prop1 & 8011 & 10460 & 1536 & 1202 \\
		
		3 & berek & 43 & 0 & 16 & 0 & 36 & prop2 & 12115 & 10899 & 1503 & 928 \\
		
		4 & camel1.0 & 327 & 12 & 13 & 0 & 37 & prop3 & 3189 & 7085 & 298 & 882 \\
		
		5 & camel1.2 & 558 & 50 & 205 & 11 & 38 & prop4 & 3384 & 5334 & 419 & 421 \\
		
		6 & camel1.4 & 802 & 70 & 144 & 1 & 39 & prop5 & 3368 & 5148 & 561 & 738 \\
		
		7 & camel1.6 & 878 & 87 & 181 & 7 & 40 & prop6 & 377 & 283 & 32 & 34 \\
		
		8 & ckjm & 10 & 0 & 5 & 0 & 41 & redaktor & 169 & 7 & 25 & 2 \\
		
		9 & elearning & 57 & 7 & 5 & 0 & 42 & serapion & 44 & 1 & 9 & 0 \\
		
		10 & forrest0.6 & 6 & 0 & 1 & 0 & 43 & skarbonka & 45 & 0 & 9 & 0 \\
		
		11 & forrest0.7 & 28 & 1 & 5 & 0 & 44 & sklebagd & 20 & 0 & 12 & 0 \\
		
		12 & forrest0.8 & 31 & 1 & 2 & 0 & 45 & synapse1.0 & 153 & 4 & 16 & 0 \\
		
		13 & intercafe & 27 & 0 & 4 & 0 & 46 & synapse1.1 & 213 & 9 & 59 & 1 \\
		
		14 & ivy1.1 & 107 & 4 & 62 & 1 & 47 & synapse1.2 & 245 & 11 & 86 & 0 \\
		
		15 & ivy1.4 & 236 & 5 & 16 & 0 & 48 & systemdata & 63 & 2 & 9 & 0 \\
		
		16 & ivy2.0 & 345 & 7 & 40 & 0 & 49 & szybkafucha & 25 & 0 & 14 & 0 \\
		
		17 & jedit3.2 & 268 & 4 & 90 & 0 & 50 & termoproject & 41 & 1 & 13 & 0 \\
		
		18 & jedit4.0 & 301 & 5 & 74 & 1 & 51 & tomcat & 796 & 62 & 77 & 0 \\
		
		19 & jedit4.1 & 308 & 4 & 79 & 0 & 52 & velocity1.4 & 179 & 17 & 132 & 15 \\
		
		20 & jedit4.2 & 363 & 4 & 48 & 0 & 53 & velocity1.5 & 198 & 16 & 133 & 9 \\
		
		21 & jedit4.3 & 474 & 18 & 7 & 4 & 54 & velocity1.6 & 211 & 18 & 76 & 2 \\
		
		22 & kalkulator & 25 & 2 & 5 & 1 & 55 & workflow & 38 & 1 & 20 & 0 \\
		
		23 & log4j1.0 & 135 & 0 & 34 & 0 & 56 & wspomaganiepi & 18 & 0 & 12 & 0 \\
		
		24 & log4j1.1 & 109 & 0 & 37 & 0 & 57 & xalan2.4 & 694 & 29 & 110 & 0 \\
		
		25 & log4j1.2 & 202 & 3 & 187 & 2 & 58 & xalan2.5 & 740 & 63 & 363 & 24 \\
		
		26 & lucene2.0 & 191 & 4 & 90 & 1 & 59 & xalan2.6 & 724 & 161 & 322 & 89 \\
		
		27 & lucene2.2 & 239 & 8 & 139 & 5 & 60 & xalan2.7 & 740 & 169 & 732 & 166 \\
		
		28 & lucene2.4 & 336 & 4 & 199 & 4 & 61 & xerces1.2 & 344 & 96 & 58 & 13 \\
		
		29 & nieruchomosci & 26 & 1 & 10 & 0 & 62 & xerces1.3 & 362 & 91 & 68 & 1 \\
		
		30 & pdftranslator & 33 & 0 & 15 & 0 & 63 & xerces1.4 & 486 & 102 & 376 & 61 \\
		
		31 & poi1.5 & 203 & 34 & 122 & 19 & 64 & xercesinit & 146 & 16 & 65 & 12 \\
		
		32 & poi2.0 & 282 & 32 & 35 & 2 & 65 & zuzel & 27 & 2 & 12 & 1 \\
		
		33 & poi2.5 & 350 & 35 & 221 & 27 &   &   &   &   &   &   \\
		\hline
	\end{tabular}  
\end{table*}

From Table \ref{NewData}, we observe that there are 35 datasets with the number of deleted defective cases greater than or equal to 1. Particularly, prop1 is the dataset with biggest number of deleted defective cases 1202. Furthermore, 54 datasets occur with the number of deleted cases greater than or equal to 1. The biggest number of deleted cases is 10899 with the dataset prop2. All in all, many datasets have shown apparent difference in the number of cases from their original datasets, which indicates that there may be performance differences when employing these two datasets for CPDP. Therefore comparative experiments will be conducted to validate whether there are prediction performance differences when using the original and cleaned Jureczko datasets for cross project defect prediction.    

\section{Comparative Experiments and Results} \label{ExperimentSection}

For the purpose of validating the essentiality of removing the inconsistent and duplicate cases in Jureczko datasets, the comparative experiments are conducted between the original Jureczko datasets and the cleaned Jureczko datasets with three training data selection approaches. Moreover, in the comparative experiments, three popular classification algorithms as well as two widely used performance evaluation metrics are employed. In the following, we will firstly introduce the three training data selection approaches. Then the experimental setup including classification algorithms and performance evaluation metrics are briefly presented. Finally the experimental results and corresponding analysis are provided detailedly.

\subsection{Training Data Selection Approaches}

In present study, three different training data selection approaches including Global Filter \cite{Menzies2013Local}, Burak Filter \cite{Turhan2009relative} and Peters Filter \cite{Peters2013Better}, are selected and will be detailedly introduced later.   

\textbf{Global Filter (GlobalF):} This filter method is described by Menzies et al. in \cite{Menzies2013Local} and Yi et al. \cite{Yi2017Training} find GlobalF performs better than eight other relevancy filters in the context of CPDP. GlobalF uses all the external project data as the training data to build the CPDP model, which means that GlobalF filter out none of any source project data.

\textbf{Burak Filter (BurakF):} This filter is a popular kind of testing case guided selection method, proposed by Turhan et al. in \cite{Turhan2009relative}. BurakF firstly uses Euclidean distance to find 10 nearest neighbors in the source data for each testing case in the target data. Then all the neighbors are combined to form a new training dataset. Finally, the new training dataset is employed to build a defect prediction model to classify the target data.  

\textbf{Peters Filter (PetersF):} This filter is proposed in \cite{Peters2013Better} by Peters et al. and it is a representative kind of training case guided selection method. Firstly, PetersF combines all the source data and target data to a combined dataset. Then the k-means clustering algorithm is used to obtain different clusters. After that, the clusters containing at least one test case are retained and each training case in the retained clusters is labelled with the nearest test case in the same cluster. Finally, for each test case, the Euclidean distance is used to select the nearest training case in the training cases that are labelled with the corresponding test case. All the selected training cases are combined into the training data for building a prediction model.

\subsection{Experimental Setup}

In this paper, three popular classification algorithms including Naive Bayes \cite{John1995estimating}, C4.5 \cite{quinlan2014c4} and Random Forest \cite{breiman2001random} have been selected. These three algorithms are reported as the most frequently used classifiers in software defect prediction \cite{Malhotra2015systematic} and we have use their Weka \cite{hall2009weka} implementations with the default values. 

We have selected F-Measure and AUC as our performance evaluation metrics in the comparative experiment. Both metrics are two of the most commonly used performance evaluation metrics in CPDP \cite{Hosseini2017A}. F-Measure is the harmonic mean between recall and precision. AUC estimates the area under the ROC curve, which is defined using the Receiver Operating Characteristic (ROC). 

\subsection{Experimental Results}

Table \ref{F-Measure} and Table \ref{AUC} show the rate of performance change of comparative experiments in terms of F-Measure and AUC, respectively. Particularly, the detailed performance change results for the combination of three classification algorithms and three training data selection methods are provided in Table \ref{F-Measure} and Table \ref{AUC}.

\begin{table*}[!htbp]
	\caption{Rate of performance change (\%) in terms of F-Measure}
	\label{F-Measure}
	\footnotesize	
	\begin{tabular}{cccccccccc}
		\toprule
		\multirow{2}{*}{Data} & \multicolumn{ 3}{c}{Naive Bayes} & \multicolumn{ 3}{c}{C4.5} & \multicolumn{ 3}{c}{Random Forest} \\
		\cmidrule(lr){2-4} \cmidrule(lr){5-7} \cmidrule(lr){8-10}
		\multicolumn{1}{c}{}&GlobalF &  BurakF & PetersF &GlobalF &  BurakF & PetersF &GlobalF &  BurakF & PetersF\\	
		\hline
		ant1.7 & 0.39  & 0  & 5.39  & -1.59  & 0  & -7.90  & -7.43  & -14  & -26.69  \\
		
		arc & 4.44  & 3.28  & 21.21  & -10.59  & 21.57  & -8.33  & 50  & 22.86  & 3.92  \\
		
		berek & 0  & 0  & -14.53  & 3.85  & 0  & -18.88  & -3.70  & 0  & 0  \\
		
		camel1.0 & 0  & 0  & -12.35  & 33.33  & 18  & 44.87  & 0  & -52  & -29.47  \\
		
		camel1.2 & 4.64  & 3.96  & 18.04  & -3.06  & -1.21  & 14.21  & -8.86  & 19.96  & 87.49  \\
		
		camel1.4 & 1.01  & 0.85  & -0.15  & -0.78  & 0.18  & 5.12  & 47.45  & -5.33  & 41.27  \\
		
		camel1.6 & 3.14  & 5.34  & -1.72  & 8.73  & 2.11  & 9.71  & -13.72  & 6.17  & 5.90  \\
		
		ckjm & 0  & 0  & 0  & 0  & 0  & 0  & 0  & 0  & 0  \\
		
		elearning & 0  & 0  & 56  & 40  & 0  & 68.42  & 0  & 0  & -100  \\
		
		forrest0.6 & 0  & 0  & 0  & 0  & 0  & 0  & 0  & 0  & 0  \\
		
		forrest0.7 & 0  & 0  & 20  & -12.50  & 0  & -7.69  & 0  & 0  & -100  \\
		
		forrest0.8 & 0  & 0  & 0  & 0  & 0  & 0  & 0  & 0  & 0  \\
		
		intercafe & 0  & 0  & 11.11  & 24.44  & 0  & -10  & 35  & 0  & -1.92  \\
		
		ivy1.1 & 1.43  & 2.67  & 4.31  & 0  & 1.30  & 9.33  & -39.22  & 29.87  & -2.47  \\
		
		ivy1.4 & 0  & 0  & 12.38  & 0  & 0.00  & 91.67  & 60.71  & 106.45  & 9.76  \\
		
		ivy2.0 & 0  & -1.01  & 0  & 2.30  & 0.00  & -2  & 63.38  & -2.50  & 23.25  \\
		
		jedit3.2 & 0  & 0  & -11.87  & 0  & 1  & 11.24  & 0  & 8  & 25.44  \\
		
		jedit4.0 & 0.81  & 0.85  & 1.87  & 1.34  & 3.45  & -6.55  & -2.96  & 0.96  & -25.61  \\
		
		jedit4.1 & 0  & 0  & -16.67  & -0.61  & 1  & -21.18  & 12.30  & -7  & -10.96  \\
		
		jedit4.2 & 0  & 0  & -1.03  & 0  & 0  & 1.47  & 19.45  & -17  & 1.10  \\
		
		jedit4.3 & 4.26  & 3.96  & -10.65  & 2.17  & 2.88  & -38.62  & -22.53  & -7.14  & -10.56  \\
		
		kalkulator & 11.11  & 10  & -100  & 0  & 0  & -100  & 0  & 0  & -100  \\
		
		log4j1.0 & 0  & 0  & -32.92  & -3.33  & 7  & -6.98  & -23.12  & -3  & -60.98  \\
		
		log4j1.1 & 0  & 0  & -39.23  & 16.23  & 24  & -30.61  & 7.32  & -1  & -57.39  \\
		
		log4j1.2 & 1.01  & -2.98  & -22.67  & 8.60  & 0.87  & -19.73  & -12.57  & -2.71  & 24.38  \\
		
		lucene2.0 & 1.00  & 0.83  & 56.07  & 0  & 19.03  & -7.27  & -29.41  & -45.28  & -32.57  \\
		
		lucene2.2 & 3.25  & 2.86  & 9.59  & 2.26  & 24.39  & 141.91  & -1.82  & 147.65  & 8.89  \\
		
		lucene2.4 & 1.68  & 1.57  & -0.40  & -1.76  & 1.51  & -17.77  & -25.44  & 25.53  & 18.23  \\
		
		nieruchomosci & 0  & 0  & 0  & 69.23  & 0  & 0  & -100  & 0  & 0  \\
		
		pdftranslator & 0  & 0  & 4.76  & 0  & 0  & 5  & 0  & 51  & -52.94  \\
		
		poi1.5 & 13.29  & 12.50  & 12.58  & 10.40  & 17.98  & 117.83  & 35.76  & -0.65  & 105  \\
		
		poi2.0 & 2.86  & 2.86  & 5.97  & -6.67  & 53.50  & 12.24  & 35.21  & -9.91  & 38.18  \\
		
		poi2.5 & 10.19  & 10.23  & 12.49  & 4.16  & -12.50  & 151.48  & 32.37  & -9.55  & -26.74  \\
		
		poi3.0 & 8.52  & 8.55  & 42.95  & 7.77  & 7.88  & 59.68  & 15.42  & -13.29  & 20.46  \\
		
		prop1 & 33.67  & 30.40  & 32.25  & 33.75  & 28.49  & 26.59  & 34.47  & 27.72  & 26.89  \\
		
		prop2 & 16.13  & 13.36  & 14.73  & 16.03  & 14.32  & 43.09  & 11.68  & 30.45  & 59.96  \\
		
		prop3 & 28.50  & 12.26  & 0.61  & 11.55  & 21.58  & 6.68  & 65.50  & 72.41  & 18.93  \\
		
		prop4 & 31.79  & 29.09  & 23.96  & 30.23  & 16.70  & 119.10  & 37.29  & 33.96  & 41.12  \\
		
		prop5 & 38.73  & 27.56  & 3.97  & 46.62  & 34.35  & 205.95  & 53.68  & 56.24  & 7.36  \\
		
		prop6 & 52.37  & -9.48  & 10.69  & 1.18  & -19.44  & 21.43  & 73.58  & 69.41  & 37.50  \\
		
		redaktor & 5.41  & 0  & -13.16  & 3.03  & 0  & 0  & -11.76  & -33  & -2.27  \\
		
		serapion & 0  & -6.25  & -100  & 15.38  & 0  & 0  & 18.18  & -100  & -100  \\
		
		skarbonka & 0  & 0  & 0  & 7.69  & 0  & -32  & 62.50  & 6  & -33.33  \\
		
		sklebagd & 0  & 0  & 26  & 0  & 0  & 0  & -39.29  & 0  & 0  \\
		
		synapse1.0 & 0  & 9.69  & -22.86  & 2.56  & 0  & -100  & 90.91  & 0  & -100  \\
		
		synapse1.1 & 1.22  & 0.97  & 47.56  & -3.50  & 9.20  & 0  & 83.56  & -29.09  & -49.28  \\
		
		synapse1.2 & 0  & 0  & -9.45  & 0.86  & 31  & -21.84  & -18.43  & -6  & -44.14  \\
		
		systemdata & 0  & 0  & -26.28  & -13.64  & 0  & -14.29  & 25.93  & 0  & -37.50  \\
		
		szybkafucha & 0  & 0  & 0  & -5.56  & 0  & 0  & 0  & 0  & 0  \\
		
		termoproject & 0  & 0  & -3.51  & 0  & 0  & -44.54  & 41.18  & 7  & -20.31  \\
		
		tomcat & 0.45  & 0  & -23.69  & -2.20  & 11  & 38.58  & -29.09  & -29  & 103.18  \\
		
		velocity1.4 & 10.27  & 9.62  & 6.50  & 16.62  & 9.74  & 45.63  & -26.48  & 30.63  & -18.23  \\
		
		velocity1.5 & 6.00  & 5.84  & 6.67  & 5.10  & -0.57  & 14.32  & -1.87  & -15.74  & 21.93  \\
		
		velocity1.6 & 2.11  & 1.94  & 6.08  & 14.25  & -22.14  & 4.76  & -8.16  & 21.15  & 15.84  \\
		
		workflow & 0  & 0  & 0  & 0  & 0  & 152  & -65.15  & 0  & 32  \\
		
		wspomaganiepi & 0  & 0  & 355  & 0  & 0  & -61.54  & 40  & 0  & 0  \\
		
		xalan2.4 & 0  & 1.39  & 3.39  & 0.41  & 2.06  & 13.16  & 11.58  & -4.25  & -12.05  \\
		
		xalan2.5 & 8.11  & 5.86  & -9.39  & 4.75  & -9.03  & -15.50  & 12.85  & -16.75  & -6.40  \\
		
		xalan2.6 & -17.07  & -0.78  & -3.61  & 20.66  & -1.23  & 45.98  & 17.83  & 24.02  & 43.17  \\
		
		xalan2.7 & -9.26  & 2.03  & 9.38  & 10.31  & 9.89  & 40.00  & 2.51  & 53.44  & 19.88  \\
		
		xerces1.2 & 12.04  & 11.50  & 10.61  & 13.51  & 15.32  & 3.17  & -11.02  & -3.33  & -42.77  \\
		
		xerces1.3 & 0.85  & 0.83  & 6.15  & 1.56  & -16.09  & 22.44  & 69.07  & -10.12  & 72.58  \\
		
		xerces1.4 & 14.02  & 13.44  & 6.95  & 9.35  & 5.62  & 109.19  & 79.91  & 38.98  & 62.12  \\
		
		xercesinit & 13.48  & 12.63  & 3.14  & 12.90  & 13.48  & 37.36  & 97.19  & 11.34  & 22.61  \\
		
		zuzel & 5.56  & 4.76  & 0  & 5.88  & -8.33  & 0  & 41.67  & -19.05  & 0  \\
		\hline
		\textbf{AVG} & 4.06  & 2.92  & 5.32  & 5.37  & 5.89  & 4.33  & 19.86  & 3.38  & -2  \\
		\bottomrule
	\end{tabular}  
\end{table*}
From Table \ref{F-Measure}, it could be observed that average rate of F-Measure change ranges from -2\% (Random Forest with PetersF) to 19.86\% (Random Forest with GlobalF). Particularly, only Random Forest with PetersF obtains the reduced prediction performance with the cleaned Jureczko datasets. In addition, we could observe from Table \ref{F-Measure} that there are performance differences for most employed datasets and many may be very great. This indicates that when using F-Measure as the evaluation metric, the performance differences between the original and cleaned Jureczko datasets are apparent.    

\begin{table*}
	\caption{Rate of performance change (\%) in terms of AUC}
	\label{AUC}
	\footnotesize	
	\begin{tabular}{cccccccccc}
		\toprule
		\multirow{2}{*}{Data} & \multicolumn{3}{c}{Naive Bayes} & \multicolumn{ 3}{c}{C4.5} & \multicolumn{ 3}{c}{Random Forest} \\
		\cmidrule(lr){2-4} \cmidrule(lr){5-7} \cmidrule(lr){8-10}
		\multicolumn{ 1}{c}{} & GlobalF &  BurakF & PetersF &GlobalF &  BurakF & PetersF &GlobalF &  BurakF & PetersF \\
		\hline
		ant1.7 & -0.75  & -0.81  & 1.91  & -1.51  & 0  & 2.53  & -7.82  & -3  & -1.14  \\
		
		arc & 2.20  & 3.64  & 2.87  & -3.38  & 0.73  & -4.11  & 19.20  & 4.62  & -4.56  \\
		
		berek & 0.26  & 0  & -9.45  & 5.06  & 0  & 3.11  & 0.41  & 0  & 0.13  \\
		
		camel1.0 & -0.81  & -1.14  & -18.70  & -6.21  & 6  & -20.84  & 13.74  & -13  & -11.86  \\
		
		camel1.2 & -0.41  & -2.60  & 1.86  & 2.43  & -4.23  & -4.03  & -0.20  & 2.64  & 2.29  \\
		
		camel1.4 & -2.48  & -3.61  & -8.63  & -0.39  & -3.06  & 4.97  & -3.49  & -7.98  & -2.60  \\
		
		camel1.6 & -0.89  & -2.52  & -2.59  & 2.72  & -1.79  & 8.35  & -1.68  & -1.68  & -6.98  \\
		
		ckjm & 0  & 0  & 0  & 0  & 0  & 0.00  & 28.57  & 0  & 0  \\
		
		elearning & -0.90  & -1.54  & 6.79  & 39.22  & -11  & 4.04  & -13.71  & 6  & -3.91  \\
		
		forrest0.6 & 0  & 0  & 0  & 0  & 0  & 0.00  & -33.33  & 0  & 0  \\
		
		forrest0.7 & -1.65  & -1.09  & 21.33  & 24.28  & 0  & 14.34  & 2.73  & 8  & -37.79  \\
		
		forrest0.8 & -4.31  & -0.69  & 0  & -4.41  & 46  & 0.00  & -6.80  & 57  & 0  \\
		
		intercafe & 0  & 0  & -17.11  & 16.08  & 0  & -3.17  & -9.03  & 0  & -8.77  \\
		
		ivy1.1 & -0.10  & 0.10  & 8.30  & -1.18  & -0.72  & -6.10  & 9.04  & 0.72  & 4.81  \\
		
		ivy1.4 & -0.38  & -0.39  & 3.72  & -0.94  & -0.51  & 32.21  & -3.21  & 6.24  & 11.63  \\
		
		ivy2.0 & -0.53  & -0.46  & 3.61  & 0.18  & -0.55  & 7.26  & 2.56  & -9.24  & 6.21  \\
		
		jedit3.2 & -0.42  & -0.62  & -2.98  & 0.64  & 0  & 0.56  & -2.35  & 0  & -0.48  \\
		
		jedit4.0 & 0.93  & 0.97  & -0.05  & 1.06  & -0.88  & -0.04  & -5.84  & 7.28  & 1.40  \\
		
		jedit4.1 & -0.22  & -0.28  & -4.78  & -0.62  & 1  & 16.96  & -0.69  & -2  & -4.91  \\
		
		jedit4.2 & -0.17  & -0.22  & 0.08  & -0.24  & 0  & 1.32  & 5.14  & -2  & -3.65  \\
		
		jedit4.3 & 39.33  & 45.97  & 47.60  & 35.27  & 16.31  & -21.31  & 48.96  & 45.16  & 60.10  \\
		
		kalkulator & 15.88  & 7.37  & -74.93  & -7.98  & 0.80  & -1.43  & 152  & 10.66  & -50.36  \\
		
		log4j1.0 & 0.26  & -0.18  & -31.50  & -2.57  & 2  & -30.72  & -3.03  & -1  & -9.96  \\
		
		log4j1.1 & 0  & -0.09  & -14.59  & -5  & 7  & -22.13  & 2.16  & -2  & 6.40  \\
		
		log4j1.2 & -2.97  & -2.10  & 1.28  & 2.11  & 5.04  & -1.96  & -4.46  & 14.16  & 22.13  \\
		
		lucene2.0 & 0.21  & 0.23  & 2.91  & -1.10  & -2.25  & -4.75  & -3.34  & -11.54  & 1.02  \\
		
		lucene2.2 & 1.27  & 0.52  & 4.50  & 0.08  & -0.91  & 11.52  & 4.55  & 2.62  & 2.12  \\
		
		lucene2.4 & 1.17  & 0.37  & -6.65  & 0.24  & 0.87  & -8.20  & 1.31  & 7.72  & -1.37  \\
		
		nieruchomosci & -0.60  & 0  & 0  & -2.12  & 0  & 0.00  & 31.44  & 0  & 0  \\
		
		pdftranslator & 0.58  & 0.51  & 27.40  & 1.52  & 0  & -4.66  & 8.21  & 7  & -4.42  \\
		
		poi1.5 & 3.67  & 2.15  & 2.93  & -6.60  & -6.30  & -3.46  & 8.87  & -4.64  & 11.30  \\
		
		poi2.0 & -0.08  & -0.02  & 13.90  & 1.32  & 16.31  & -2.89  & -1.72  & 9.70  & 20.80  \\
		
		poi2.5 & 2.21  & -0.28  & -7.35  & -5.40  & -6.46  & 9.61  & 6.78  & -2.58  & 3.52  \\
		
		poi3.0 & 0.37  & -1.08  & -5.56  & 12.37  & 0.93  & 28.84  & 8.65  & 8.75  & 2.04  \\
		
		prop1 & 7.85  & 8.84  & 10.60  & 1.71  & 3.95  & 4.15  & 3.39  & 4.91  & 3.36  \\
		
		prop2 & 3.92  & 3.78  & 4.40  & -0.82  & -6.32  & -1.71  & -0.48  & 8.86  & 11.06  \\
		
		prop3 & 9.65  & 9.90  & 4.51  & 4.32  & 15.59  & 10.76  & 10.12  & 14.61  & 7.43  \\
		
		prop4 & 10.45  & 11.33  & 8.98  & 3.98  & -3.48  & 11.36  & 8.07  & 0.17  & 12.32  \\
		
		prop5 & 7.09  & 5.04  & 6.06  & 3.30  & 3.95  & 12.48  & 6.03  & 7.43  & -0.44  \\
		
		prop6 & 6.14  & 4.56  & 6.65  & 3.09  & 13.97  & 4.19  & 11.53  & 9.25  & 9.05  \\
		
		redaktor & 4.16  & 3.63  & 4.37  & -1.09  & -7  & -26.31  & 12.77  & -11  & 10.12  \\
		
		serapion & -0.83  & 6.05  & -8.69  & 7.87  & 0.15  & 28.57  & -1.92  & -4.60  & 55.25  \\
		
		skarbonka & 0  & 0  & 12.22  & 1.91  & 0  & -17.49  & -6.25  & -5  & 4.64  \\
		
		sklebagd & 0  & 0  & -9.38  & 0  & 0  & 0.00  & 2.84  & 0  & 104.55  \\
		
		synapse1.0 & -0.65  & -0.37  & -33.32  & -0.38  & 0.15  & -20.90  & 9.03  & 11.09  & -18.69  \\
		
		synapse1.1 & -0.28  & -0.37  & 14.54  & -2.20  & -10.05  & 0.00  & 11.82  & 5.38  & 7.87  \\
		
		synapse1.2 & -2.11  & -1.84  & 9.39  & -0.84  & 0  & -2.99  & -2.43  & -5  & -12.48  \\
		
		systemdata & 1.04  & -0.01  & -3.96  & -20.77  & -1  & -1.76  & -11.59  & 20  & -5.28  \\
		
		szybkafucha & 0  & 0  & -13.46  & -6.73  & 0  & -9.09  & -24.90  & 0  & 32.73  \\
		
		termoproject & -1.09  & -0.25  & -20.52  & 2.47  & 0  & -38.87  & -9.81  & 10  & 2.03  \\
		
		tomcat & -1.73  & -1.59  & -8.35  & -7.03  & 1  & -19.18  & -8.26  & -6  & 4.55  \\
		
		velocity1.4 & 5.26  & 3.21  & 85.79  & -6.10  & -2.43  & -17.69  & -9.84  & -0.02  & 33.63  \\
		
		velocity1.5 & 1.72  & -0.82  & 1.44  & -5.79  & 6.93  & 8.15  & -7.55  & -1.33  & 15.79  \\
		
		velocity1.6 & -1.53  & -2.85  & -12.61  & 1.75  & -3.04  & 15.07  & 3.37  & 0.18  & -6.01  \\
		
		workflow & -3.24  & -2.64  & -12.45  & 11.63  & -0.26  & -5.05  & -3.68  & -2.55  & -14.40  \\
		
		wspomaganiepi & 0  & 0  & 0  & 4.95  & 0  & -13.33  & 9.09  & 0.00  & 23.58  \\
		
		xalan2.4 & -1.03  & -1.58  & -0.47  & -1.62  & -0.88  & -11.83  & 6.71  & 3.17  & -4.96  \\
		
		xalan2.5 & 1.34  & 0.87  & -1.48  & 1.36  & -2.90  & -5.77  & -4.13  & -2.08  & -10.09  \\
		
		xalan2.6 & -4.26  & 1.99  & -3.41  & 23.55  & -5.42  & 6.51  & 10.88  & 1.88  & 3.47  \\
		
		xalan2.7 & -4.92  & -2.14  & -0.99  & 3.44  & 3.07  & -0.65  & -17.19  & -1.69  & 12.02  \\
		
		xerces1.2 & -3.48  & 0.50  & 4.69  & 0.78  & 0.47  & -6.94  & 0.45  & 25.27  & -1.37  \\
		
		xerces1.3 & -6.19  & -6.16  & -8.44  & -0.82  & 19.83  & 26.34  & 4.76  & -5.90  & -6.46  \\
		
		xerces1.4 & 1.25  & -1.79  & -3.37  & -0.52  & 4.49  & 1.25  & -0.19  & -2.72  & 3.40  \\
		
		xercesinit & 5.70  & 9.05  & 3.53  & 1.58  & 1.93  & 15.91  & 3.71  & 17.08  & 7.91  \\
		
		zuzel & 5.71  & 8.02  & 0  & 3.60  & -0.85  & 0.00  & 8.26  & 12.29  & 0  \\
		\hline
		\textbf{AVG} & 1.19  & 1.17  & -2.32  & 1.55  & 1.43  & -3.02  & 2.28  & 3.09  & 0.38  \\
		\bottomrule
	\end{tabular}  
\end{table*}
From Table \ref{AUC}, it could be observed that the average ratio of AUC change ranges on a small scale compared with the average ratio of F-Measure change in Table \ref{F-Measure}. In fact, the average ratio of AUC change ranges from -3.02\% (C4.5 with PetersF) to 3.09\% (Random Forest with BurakF). However, compared with Table \ref{F-Measure}, we observe that more datasets obtain the performance difference as there are less 0 in Table \ref{AUC} than in Table \ref{F-Measure}. 

In conclusion, from Table \ref{F-Measure} and Table \ref{AUC}, it could be found that most datasets have obtained the different defect prediction  performance when using the original and cleaned Jureczko datasets separately. This means that the quality problems of identical and inconsistent cases in these datasets indeed influence the performance of cross project defect prediction, which indicates that for the purpose of obtaining more actual and reliable prediction performance for CPDP, it is necessary to deal with the problematic cases in the original Jureczko datasets. 
   
\section{Conclusion}
Cross project defect prediction has been widely studied in recent decade. In the research of CPDP, Jureczko and NASA datasets are two most widely used public datasets. Researchers have found the importance of selecting appropriate training data for building a CPDP model. However, though the NASA datasets have been reported to have data quality problems, there is still no research focusing on the data quality problems of Jureczko. Therefore in present study, we have firstly investigated the data quality problems in the Jureczko datasets and found that there are many datasets with identical cases and inconsistent cases. Then a data cleaning method is proposed to deal with the original Jureczko datasets and a cleaned version of Jureczko datasets is provided. Finally, the comparative experiments are conducted with the original and cleaned Jureczko datasets. In the comparative experiments, three training data selection methods and two classification algorithms are employed with using F-Measure and AUC as the performance evaluation metrics. The experimental results show that many problematic datasets obtain very different defect prediction performance, which indicates the effectiveness and necessity of our study of public data quality problems in cross project defect prediction.  


\bibliographystyle{ACM-Reference-Format}
\bibliography{CrossProjectDataQuality}

\end{document}